\documentclass[a4paper,english]{article}
\usepackage[T1]{fontenc}
\usepackage[latin1]{inputenc}
\usepackage{color}
\usepackage{babel}
\usepackage{url}
\usepackage{amsmath}
\usepackage{natbib}
\usepackage{amssymb}
\usepackage{graphicx}
\usepackage[unicode=true,
 bookmarks=true,bookmarksnumbered=false,bookmarksopen=false,
 breaklinks=false,pdfborder={0 0 1},backref=section,colorlinks=true]
 {hyperref}
\hypersetup{
 pdfauthor={Romain},
 pdftex,linkcolor=blue,citecolor=blue,filecolor=blue,urlcolor=blue,pdftitle=,pdfsubject=,pdfkeywords=}
\usepackage{breakurl}

\makeatletter

\providecommand{\tabularnewline}{\\}

\usepackage{babel}

\@ifundefined{showcaptionsetup}{}{%
 \PassOptionsToPackage{caption=false}{subfig}}
\usepackage{subfig}
\makeatother

\begin{document}

\title{Social Interactions vs Revisions, What is important for Promotion
in Wikipedia?}

\author{Romain Picot-Clémente\textsuperscript{1,2}, Cécile Bothorel\textsuperscript{1,2},
Nicolas Jullien\textsuperscript{1,3}\footnote{\textsuperscript{1}Institut Mines Telecom Bretagne, \textsuperscript{2}UMR
CNRS 3192 Lab-STICC, \textsuperscript{3}ICI-M@rsouin, \{Romain.PicotClemente,
Cecile.Bothorel, Nicolas.Jullien\}@telecom-bretagne.eu}}

\begin{abstract}
In epistemic community, people are said to be selected on their knowledge
contribution to the project (articles, codes, etc.) However, the socialization
process is an important factor for inclusion, sustainability as a
contributor, and promotion. Finally, what does matter to be promoted?
being a good contributor? being a good animator? knowing the boss?
We explore this question looking at the process of election for administrator
in the English Wikipedia community. We modeled the candidates according
to their revisions and/or social attributes. These attributes are
used to construct a predictive model of promotion success, based on
the candidates's past behavior, computed thanks to a random forest
algorithm.

Our model combining knowledge contribution variables and social networking
variables successfully explain 78\% of the results which is better
than the former models. It also helps to refine the criterion for
election. If the number of knowledge contributions is the most important
element, social interactions come close second to explain the election.
But being connected with the future peers (the admins) can make the
difference between success and failure, making this epistemic community
a very social community too. 
\end{abstract}
\maketitle

\section{Introduction}

Mobilizing hundreds (Linux) to thousands of contributors (Wikipedia),
volunteer online open projects aiming at creating new knowledge, online
``communities of creation\textquotedblright , as named by \citet{RullaniHaefliger13},
are viewed as central in the generation of new, innovative knowledge
by and for firms. But the path to successful community building is
still risky and uncertain, and as for business building, most of the
attempts fail, no matter how many hundreds of thousands of dollars
were put into them \citep{Worthen08}.

One of the key elements to develop a successful and sustainable community,
as explained a quarter of century ago by Eric von Hippel \citeyearpar{vonHippel86},
is to attract enough highly competent and ``committed/committing''
contributors, being they named ``lead users'', ``core'', or ``big''
contributors \citep{MahrLievens12,FangNeufeld09}, i.e. the most productive
people, who are also those with more responsibility in the management
of the project \citep{RullaniHaefliger13}. How, on what criterion
these core contributors are recruited to become managers is still
a matter of research, something we want to address here.

If we agree with the theory of epistemic community \citep{CohendetCrepletDupouet01,Edwards01},
which stresses that those communities are project-oriented communities
of experts, evaluated on their contribution in terms of knowledge,
the main criterion for promotion in the different steps of the career
will be their knowledge production. For Wikipedia, when projects have
rules for running for administrator, they are about knowing the rules,
but also about the number of edits of articles (more than 3,000 and
more of one year of activity for the French Wikipedia, \url{https://fr.wikipedia.org/wiki/Wikipédia:Candidature_au_statut_d'administrateur}).
\citet[part II]{OMahonyFerraro07}, studying Open Source projects,
showed that ``developers who were making greater technical contributions
and who were more engaged in organization building were more likely
to become members of the leadership team''. (p. 1096). \citet{FlemingWaguespack07}
found the same result in their study of the Internet Engineering Task
Force community.

However, \citet{Pentzold10}, for Wikipedia, \citet{VonKroghetal12},
for open source, defended the idea that becoming a big contributor
may be an additional step from being a regular contributor, an additional
commitment, which would occur for reasons developed during the attendance
of the project as the development of this sense of \textquotedblright community\textquotedblright ,
i.e. to understand and accept the rules of the organization \citep{Butleretal08,Cardon12}.
If we follow their argument, the social interactions with peers may
be an additional requirement for being promoted, as explained by \citep{RullaniHaefliger13}.

This article discusses whether the knowledge contribution or the social
connexion matter more for being promoted in an online epistemic community,
looking at the electing process of the administrators (admin) in the
English Wikipedia, called the Request for Adminship (RfA). Any user
participating on Wikipedia can request to become administrator, but
every candidate is not elected, as RfA has about a 4 in 10 chance
of being successful.

The article is organized as follows: in section 2 a review of the
literature used to construct our framework of investigation, in section
3 the formulation of our hypotheses, in section 4 the data collection
strategy (choice of the community and definition of the questions),
in section 5 the results. We discuss the consequences of this work,
its limits and future research in section 6 before concluding.


\section{Literature review}

The fact that these projects are made possible by the aggregation
of various motivations and levels of involvement is classic for social
theories, being the critical mass theory, which regards the construction
of collective action \citep{OliverMarwellTeixeira85,MarwellOliver93},
or the theory of (knowledge) commons \citep{Ostrom90,HessOstrom06b}.

More preciselly, \citet{Shah06} showed, in the case of open source,
that long-term participants enjoyed programming and interacting with
the rest of the community (i.e., labeled as ``hobbyists\textquotedblright ),
whereas short-term participants were typically driven by an immediate
need for software (i.e., use value). Theoretical analyses of incentives,
in software projects \citep{ForayZimmermann01,LernerTirole02} or
in wikis (\citet{ForteBruckman05}, using \citeauthor{LatourWoolgar79}'s
(1979) analysis of science ``cycles of credit''), estimate that
the other main vector for participation, beside being able to have
feedbacks and improvements on the pieces of knowledge proposed, is
the quest for reputation. Applied works on Wikipedia \citep{Nov07,YangLai10,ZhangZhu11},
professional electronic networks \citep{WaskoFarah05,JullienRoudautleSquin11},
and open source software \citep{Shah06,Scacchi07}, confirm that peer
recognition, whether it be professional or community recognition,
is a main motive for contribution amongst the main contributors, in
addition to intrinsic factors (personal enjoyment and satisfaction
from helping by sharing their knowledge).

In the same time, the role of a person evolves over time in the communities
of creation \citep{VonKroghetal03,JensenScacchi07}, as do the reasons
why s/he participates \citep{VonKroghetal03}. Regarding Wikipedia,
for instance, it has been showed that there is process of specialization
of the editors \citep{Ibaetal10,Welseretal11}, with multipurpose
wikipedians, able to participate in the redaction of an article, and
wikipedians focused on the global editing, i.e. the coördination and
the organization of the project.

Consequently, contributors may develop their social connections, interacting
with peers and this may be drive social recognition. Beyond a certain
level of contribution, one can even wonder if these social interactions
matter more than additional knowledge production for receiving peer
recognition. More precisely, we wonder if it is possible to predict
the promotion of a user according to her activity, separating this
activity into knowledge production (i.e. ``edits'' of the articles,
or ``revisions'', for Wikipedia, ``commits'' for open source software),
and social activity. This is what we tested here, using the English
Wikipedia as case study.

\section{Choice of the case study}

Although Open Source initiatives are numerous, in various industries
\citep{Balkaetal09}, the main open knowledge project outside the
computer industry is to be found in the encyclopedia editing project
known as Wikipedia. It has become one of the most successful knowledge
production projects ever, with more than 4 million articles for the
English version and more than one million visits per day, and is seen
as a model for knowledge management theory \citep{Mcafee06,HasanPfaff06b}.

In addition Wikipedia has a process of election for some of the managing
task, the administrator position, where social connections and knowledge
production skills seem to matter, and provides a complete set of data
regarding this process.

The Administrators have more rights than normal users on Wikipedia;
they can (un)block specific users from editing pages, they can do
some special actions on pages like (un)protecting from editing, (un)deleting,
renaming, reporting vandalism, etc. In Wikipedia, every registered
user can request to be promoted administrator. Nevertheless, not all
RfA are successful, it is not easy to become an administrator. Precise
rules for applying vary from one language to another. As already said,
we will focus on the English Wikipedia and its rules here, where the
RfA process is as follows. First, candidate creates a page dedicated
to the request. During seven days which can be assimilated to a campaign,
the candidate is questioned and his characteristics are studied by
anybody from the community. Every user can vote for or against the
candidate, and can change her mind at any moment during this period.
At the end of the campaign, following the votes and the discussions
on the RfA page, special users called ``bureaucrats'' gives their
verdict about accepting or rejecting the RfA. There is no objective
threshold on the percentage of support votes needed to be elected
by bureaucrats. Nevertheless, it appears that a candidate is more
likely to pass if he achieves at least 80\% support. With less than
70\% support, the candidate is generally not promoted. From 2001 to
2008, on the English Wikipedia, 2794 users have requested to become
administrators. Among them, 1248 requests have been accepted, so that
the success rate is about 44.7\%.

Previous studies shed light on the entanglement of the knowledge production
skills and of the social skills for being promoted. Considering the
knowledge skills, \citet{LeskovecHuttenlocherKleinberg10c} showed
that voters are more likely to give positive votes when candidates
are more active than them in terms of ``edits''. the flip side of
the coin is that they also showed that voters who have spoken to the
candidate before the RfA tend to cast a positive ballot. Still regarding
the importance of socialization, \citet{Leeetal12} considering the
implicit social network from talk pages of users showed that voter's
vote was highly correlated with the one of his neighbors in the network:
voters were more likely to participate to elections involving their
contacts; influent users participating in a vote can influence the
final result. And the probability of a positive vote, with an accuracy
of 84\%, is function of the intensity of the voter's relations (i.e.
co-editions, discussions, co-revert%
\footnote{The co-reverts acting negatively.%
}) with the candidate \citep{Jankowskietal13}.

Closer to what we want to study here, \citet{BurkeKraut08b} proposed
a model to predict RfA results according to the criteria put forward
by the Guide to RfAs%
\footnote{\url{http://en.wikipedia.org/wiki/Wikipedia:GRFA }%
}, where are described the criteria RfA evaluators look for in nominees,
mainly based on accounting candidate's activity: Strong edit history,
with Edit summaries (explaining what they did when editing), and High
quality of articles, Varied experience, User interaction, Helping
with chores (i. e. already working on admin tasks such as discussing
articles for delation), Trustworthiness, Observing consensus, and
having various experiences in terms of editing, user interaction,
etc. Their model's accuracy reached 75.6\%. However, as the authors
mentioned, their measure of some variables, the less obvious, such
as trustworthiness, if simple to compute, are quite naïve. On the
other hand, as it has been proved in the case of measuring articles'
quality, it is not sure that increasing the complexity of the measure
improves its accuracy very much%
\footnote{the best indicators of an Feature Article, at least for the English
version \citep{Dalipetal09} are the length and basic quality of the
writing, as it is for open source contribution, where \citet{HofmannRiehle09}
found that the simple heuristics are superior to the more complex
text-analysis-based algorithms to estimate the size and the importance
of a commit in open-source projects.%
}.

Moreover, regarding the question we address, here, they did not measure
the respective influence of the edits and of the social interaction
on RfA result. They did not separate social networking with administrators
from social networking with everyone either, whereas an administrator
(or a bureaucrat) may be more influent than an unknown user on an
RfA result, as showed by \citep{Leeetal12}.

\section{Data collection and model}

To measure the respective importance of the knowledge contribution
and the socialization for being promoted, we separated the profile
of a candidate in two parts. A revision part, which focuses on the
revision activities of the candidates, and a social part which is
based on their social activities. The social part is computed in two
social graphs, one considering interaction between the candidate and
every user, and one considering only the candidate and administrators.
We choose to consider two graphs because of the hypothesis that socialization
with administrators, the future ``peers'', has more influence on
the promotion than socialization with anybody.

\subsection{Dataset}

For this study, for convenience but also to be able to benchmark our
results with the previous studies we presented, we used a dataset
given by the Stanford Large Network Dataset Collection%
\footnote{\url{https://snap.stanford.edu/data/##wikipedia}%
} on Wikipedia. We focused on the RfA occurred in the period of time
from 2006-01-01 to 2007-10-01, because this period contains an important
number of RfA, and because there were sufficient activities before
2006 to construct the social networks based on user talks. In this
considered period, we removed the RfA done several times in a month
by a same user. Hence, the resulting number of RfA in the dataset
we considered was 1,617, with a success rate of 49.2\%.

\subsubsection{The variables}

In this part, we describe the different features we consider for modeling
candidates.

\paragraph{The revision part}

It is based on user's revision activities.

From these activities, we extracted the number of revisions/editions
they made (variable: $Revision$), the number of distinct pages ($Pages$)
they edited and the number of distinct categories ($Categories$)
they participated in, and finally, the repartition of their revisions
($Revision_{repartition}$$)$ in order to take into account both
the volume and the variety of the revisions%
\footnote{For this, we calculated the Gini coefficient on the number of revisions
of the user by pages. This attribute allows to quantify the inequalities
in a distribution. If it approaches 1, it means that the user was
mainly focused on few pages among the whole set of pages revised.
Inversely, if it approaches 0, it means the user has had an equal
revision behavior on every page revised.%
}.

Then, we assumed that the users's talks on the discussion pages of
the articles are related to their revision activities, and we added
three attributes about this talking on pages: the number of distinct
pages where the user talked ($TalkPages$), the total number of talks
on the articles's discussion pages ($PageTalks$), and the repartition
of the user talks on these pages ($PageTalks_{repartition}$).

\paragraph{The social part}

We focused on the conversations on the users' pages, to assess the
impact of what happens beside the discussion on the edits, and more
generally, beside the interactions regarding the production of knowledge.

We created three weighted and oriented graphs, based on the social
interaction, a general one where the nodes are the considered candidate
and all the users, named $userSN$, and two specific ones: 
\begin{itemize}
\item a graph where nodes are the considered candidate and all the (already)
admins, named $adminSN$; 
\item a graph where nodes are the considered candidate and all the (already)
bureaucrats, named $burSN$. 
\end{itemize}
For each graph, we computed the attributes that described the characteristics
of the node 'candidate'. These attributes are described bellow. As
they are the same for each graph, we only gave one name for each type
of attribute, and added a suffix which is the name of the related
graph.

The first attribute is the degree of the node ($Degree$), without
taking into account the orientation of edges. Then, for more details,
we considered 1) the out degree of the node ($outDegree$) which represents
the number of distinct users/admins/bureaus to whom the candidate
posted a message to on their user page, 2) the in degree of the node
($inDegree$) which is the number of distinct users/admins/bureaus
that posted a message on the candidate's page. Then, we considered
the total number of messages posted and received by the candidate
($TalksNumber$). It is different from the Degree since the weightings
are used here. The graph being oriented, we took also into account
the total number of messages posted to users/admins/bureaus pages
($outTalksNumber$) and the total number of messages received by the
candidate ($inTalksNumber$).Then, we computed multiple centrality
measures on the graphs: 
\begin{itemize}
\item the closeness centrality attribute ($Closeness$). This metric is
the inverse of the sum of the distances from this node to all the
other nodes. The more central a node is, the lower is its total distance
to all other nodes. It can be seen as a measure of how long it will
take to spread information from the candidate node to all other nodes
sequentially \citep{Beauchamp65}; 
\item the PageRank centrality ($PageRank$)\citep{Pageetal99}, a classic
metric on graphs which gives an indicator on whether the candidate
node is centric in the graph. 
\item the betweeness centrality ($Betweeness$) which is equal to the number
of shortest paths from all nodes to all others that pass through the
candidate's node. A node with high betweenness centrality has a large
influence on the transfer of information through the network, under
the assumption that information transfer follows the shortest paths
\citep{Freeman77}. This latter measure have not been computed on
the general graph because of limited computer capacity. 
\end{itemize}
Finally, we computed the Gini coefficient for both the number of messages
posted by the candidates $(\mathit{outTalks_{repartition}})$ and
the number of messages received by them $(\mathit{inTalks}_{\mathit{repartition}})$.
These attributes allow to quantify the repartition of the messages
from or for the candidates. As said before, a low value (0) means
a dispersion behavior whereas a high value means a focused one.

\subsection{The models}

We created multiple predictive models of RfA success based on the
random forest algorithm. This algorithm is a learning method for classification
(and regression) that operates by constructing a multitude of decision
trees \citet{Quinlan93} during training time, and outputting the
class that is the dominant value (mode) of the classes output by individual
trees. More details can be seen in \citet{Breiman01}.

\begin{figure}
\begin{centering}
\includegraphics[scale=0.3]{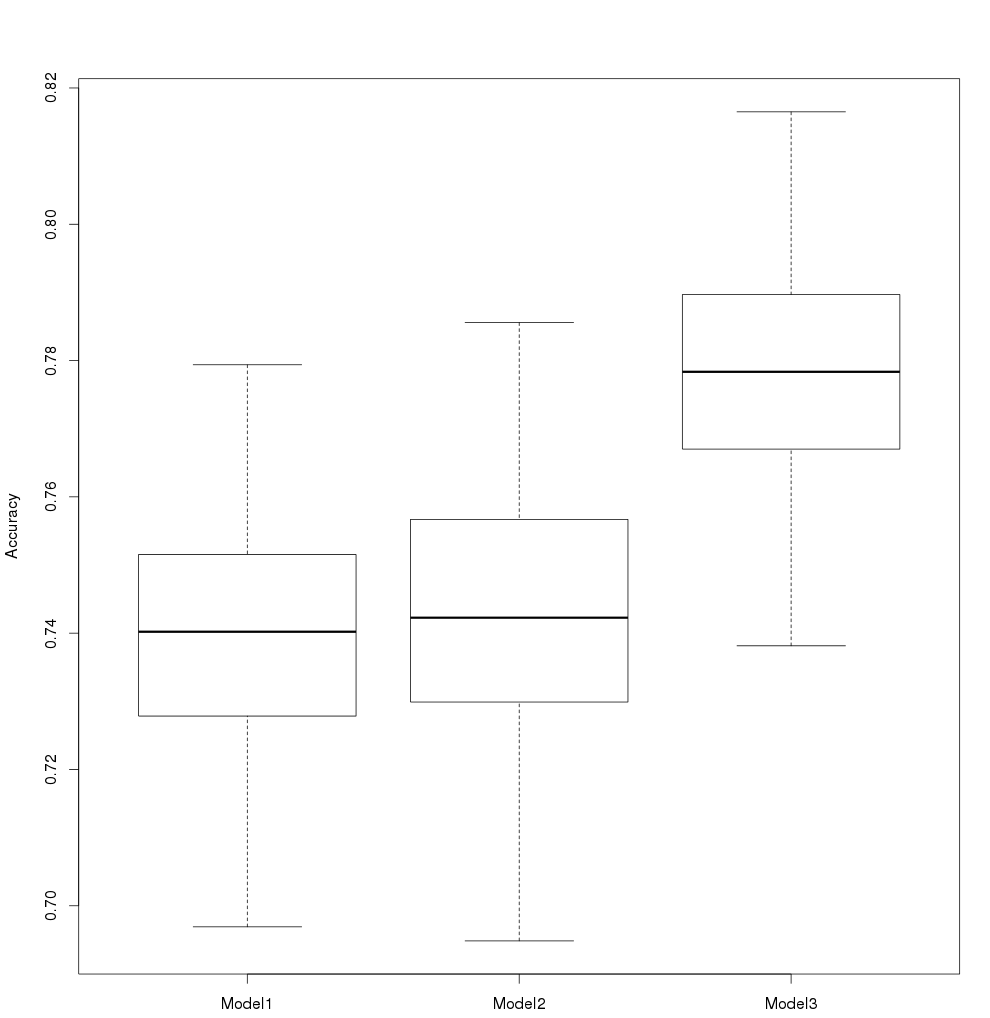} 
\par\end{centering}

\protect\protect\caption{\label{fig:Prediction-accuracy-for-1}Prediction accuracy for each
predictive model}
\end{figure}

Each predictive model considered a different modeling of candidate
profiles, taking into account subsets of features from the modeling
proposed in the previous section. Since we want to understand the
contribution of the social attributes in the RfA result, we first
created two predictive models, one based on the revision attributes
and one based on the social attributes. Then, we considered a model
using every attributes. The different types of profiles for each model
are described below: 
\begin{enumerate}
\item The profile based on the revision variables used:\\
 $\mathit{Revisions}$, $Pages$, $\mathit{Categories}$, $TalkPages$,
$PageTalks$, $\mathit{Revision}_{\mathit{repartition}}$, $\mathit{PageTalks}_{\mathit{repartition}}$ 
\item The profile based on the social variables used:\\
\begin{minipage}[t]{1\columnwidth}%
$\mathit{Degree}_{\mathit{adminSN}}$, $out\mathit{Degree}_{\mathit{adminSN}}$,
$in\mathit{Degree}_{\mathit{adminSN}}$, $TalksNumber_{\mathit{adminSN}}$,
$outTalksNumber_{\mathit{adminSN}}$, $inTalksNumber_{\mathit{adminSN}}$,
$\mathit{Closeness}_{\mathit{adminSN}}$, $\mathit{PageRank}_{\mathit{adminSN}}$,
$\mathit{Betweeness}_{\mathit{adminSN}}$, $\mathit{outTalksRepartition}_{\mathit{adminSN}}$,
$\mathit{inTalksRepartition}_{\mathit{adminSN}}$, $\mathit{Degree}_{\mathit{userSN}}$,
$out\mathit{Degree}_{\mathit{userSN}}$, $in\mathit{Degree}_{\mathit{userSN}},$
$TalksNumber_{\mathit{userSN}}$, $outTalksNumber_{\mathit{userSN}}$,
$inTalksNumber_{\mathit{userSN}}$, $\mathit{Closeness}_{\mathit{userSN}}$,
$\mathit{PageRank}_{\mathit{userSN}}$, $\mathit{outTalksRepartition}_{\mathit{userSN}}$,
$\mathit{inTalksRepartition}_{\mathit{userSN}}$, $\mathit{Degree}_{\mathit{burSN}}$,
$out\mathit{Degree}_{\mathit{burSN}}$, $in\mathit{Degree}_{\mathit{burSN}}$,
$TalksNumber_{\mathit{burSN}}$, $outTalksNumber_{\mathit{burSN}}$,
$inTalksNumber_{\mathit{burSN}}$, $\mathit{Closeness}_{\mathit{burSN}}$,
$\mathit{PageRank}_{\mathit{burSN}}$, $\mathit{Betweeness}_{\mathit{burSN}}$,
$\mathit{outTalksRepartition}_{\mathit{burSN}}$, $\mathit{inTalksRepartition}_{\mathit{burSN}}$%
\end{minipage}
\item The profile based on both the social and the revision parts, which
used the whole set of variables of the two firsts. 
\end{enumerate}
For each predictive model, we separated the dataset in a train and
a test sets. The train set consisted in a random 70\% of all the candidates
and the test set contained the remaining 30\%. Then, the predictive
model was trained on the train set and applied on the test set to
predict RfA success. We compared those predictions to the real RfA
success value of the test set, to deduce an accuracy value for each
predictive model. Accuracy is the ratio of the number of good predictions
on the number of predictions. This process was done 100 times to smooth
extreme cases. We present the boxplots of the results in Figure \ref{fig:Prediction-accuracy-for-1}.

There are many attributes in model 3 and some of them may be useless,
being very correlated to others. Hence, we calculated the Pearson
correlation on all attribute's pairs and we removed one element of
the pair for which the absolute value of the correlation is over 0.8.
More precisely, the attributes: Pages, PageTalks, Degree\_{*}SN, outTalksNumber\_userSN,
inTalksNumber\_userSN, TalksNumber\_{*}SN, inTalksNumber\_adminSN,
outTalksNumber\_adminSN, Closeness\_{*}SN, Betweeness\_{*}SN are very
correlated to other ones.

\begin{figure}
\begin{centering}
\includegraphics[scale=0.3]{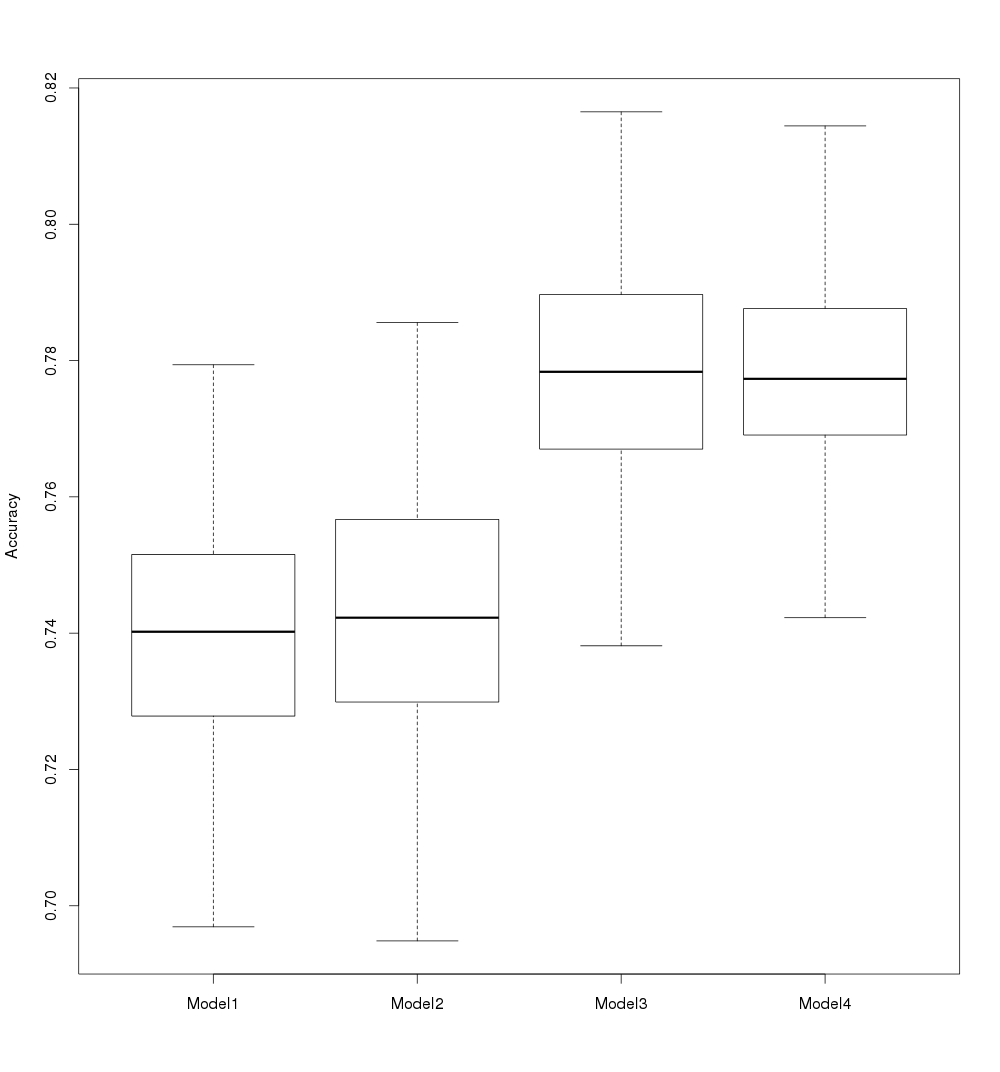} 
\par\end{centering}

\protect\protect\caption{\label{fig:with-model-4}Prediction accuracy with model 4}
\end{figure}

After this operation, we obtained a new model (Model 4) which is as
good in prediction and better in variance (Figure \ref{fig:with-model-4}).

Table \ref{tab:Confusion-matrix-for} details the accuracy of the
models in predicting either an unsuccessful (0) or a successful (1)
promotion, giving the confusion matrix for model 1, model2, and model4.

\begin{table}
\protect\protect\caption{\label{tab:Confusion-matrix-for}Confusion matrix for each predictive
model}

\centering{}%
\begin{tabular}{|c|c|c|c|c|}
\hline 
 &  & \multicolumn{3}{c|}{Confusion Matrix}\tabularnewline
\hline 
\hline 
 &  & 0  & 1  & Accuracy\tabularnewline
\hline 
Model 1  & 0  & 169  & 77.5  & 68.6\%\tabularnewline
\hline 
(Revisions)  & 1  & 48  & 190  & 79.8\%\tabularnewline
\hline 
Model 2  & 0  & 168  & 80  & 67.7\%\tabularnewline
\hline 
(Social)  & 1  & 46.5  & 192  & 80.5\%\tabularnewline
\hline 
Model 4  & 0  & 177  & 70  & 70.6\%\tabularnewline
\hline 
(Social + Revisions)  & 1  & 37  & 200  & 84.4\%\tabularnewline
\hline 
\end{tabular}
\end{table}

\section{Results}

The predictive model based on revision attributes (model 1) and the
one based on social attributes (model 2) are almost equivalent in
terms of quality of RfA results prediction : median accuracies are
respectively 74.0\% and 74.2\%. In detail, following to the confusion
matrices of model 1 and model 2, the accuracies are respectively 79.8\%
vs 80.5\% when predicting successful promotions and 68.6\% vs 67.7\%
when predicting unsuccessful ones.

When aggregating social and revisions (Model 4), median prediction
accuracy rises up to 77.8\%, while the accuracy is 84.4\% for predicting
successful promotions and 70.6\% for predicting unsuccessful ones.

According to these results, social and revisions attributes seem to
be complementary for predicting RfA results.

\begin{figure}
\begin{centering}
\subfloat[Importance of attributes for prediction accuracy ]{\begin{centering}
\includegraphics[scale=0.35]{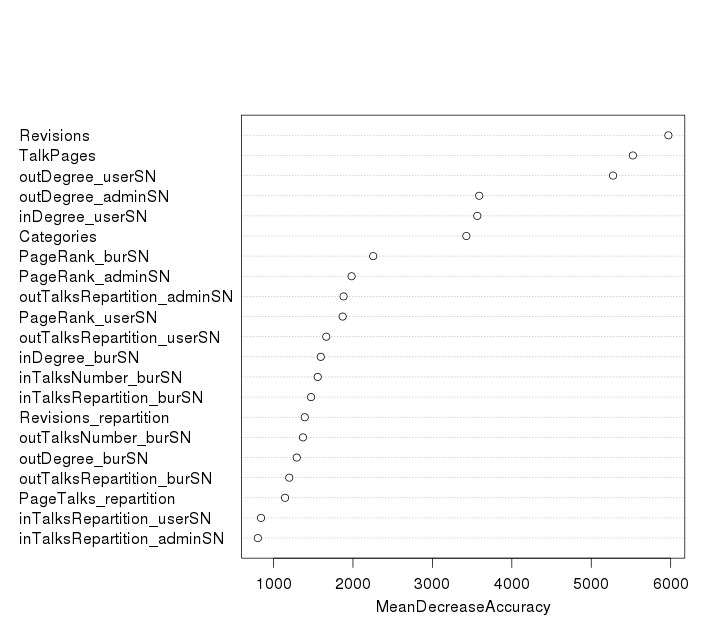} 
\par\end{centering}

}
\par\end{centering}

\centering{}\subfloat[Importance of attributes for predicting unsuccessful promotions]{\begin{centering}
\includegraphics[scale=0.18]{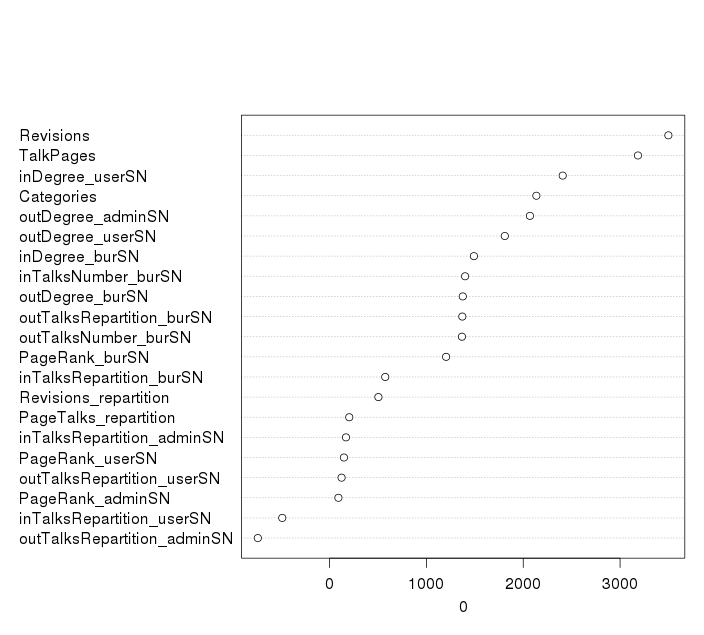} 
\par\end{centering}

}

\subfloat[Importance of attributes for predicting successful promotions]{\begin{centering}
\includegraphics[scale=0.18]{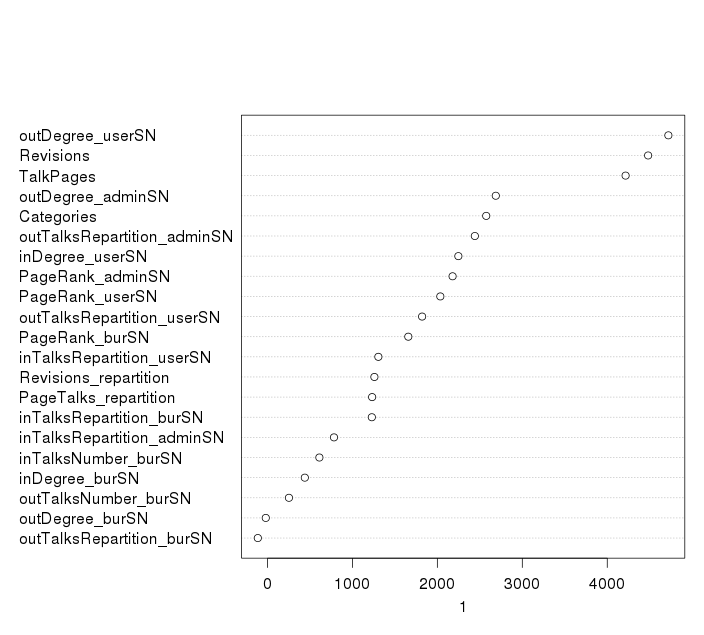} 
\par\end{centering}

}

\protect\protect\caption{\label{fig:Attribute-importance-in-1}Importance of attributes in
predictive model 4}
\end{figure}

The random forest method gives values for quantifying the importance
of an attribute for the prediction quality. For this purpose, it computes
the average decrease of accuracy of each tree into the forest when
a given attribute is not used. Higher this value is, more important
this attribute is for the prediction. Figure \ref{fig:Attribute-importance-in-1}
presents these attributes by ascending order of importance for Model
4.

In this figure, we can see that the most important attributes in general
are Revisions, TalkPages, outDegree\_userSN. In particular, to predict
the successful promotions, these three attributes are also standing
out but not in the same order: outDegree\_userSN, Revisions, TalkPages.
To predict the unsuccessful promotions, two attributes stand out:
Revisions and TalkPages.

\begin{figure}
\centering{}\includegraphics[scale=0.15]{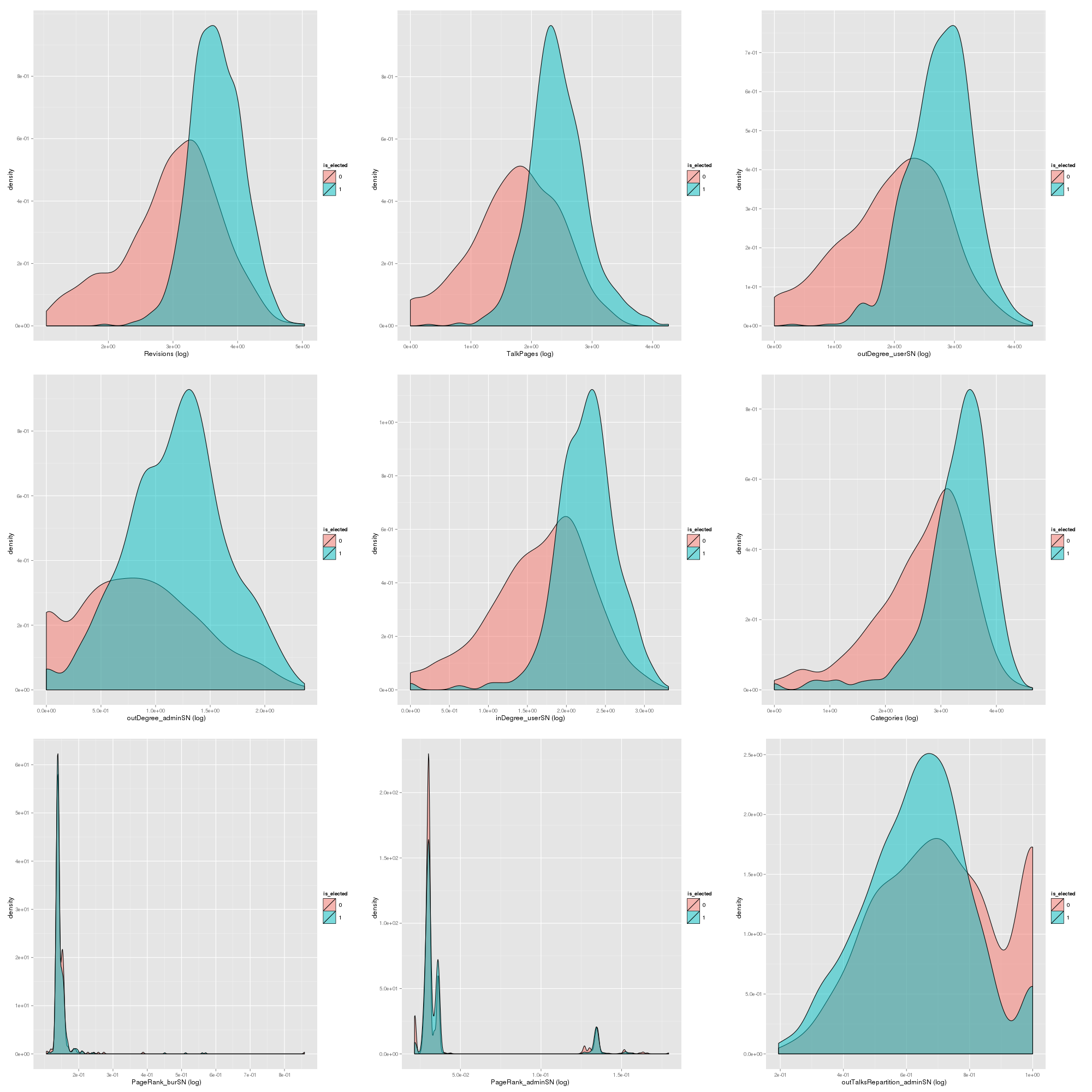}\protect\protect\caption{\label{fig:Density-of-probabilities-1}Density of probabilities of
each attribute for promoted and non-promoted candidates}
\end{figure}

These results do not give any information on the preferred values
for each attribute. For this purpose, we first compared the density
of probabilities of every attribute between accepted candidates and
rejected ones. Figure \ref{fig:Density-of-probabilities-1} shows
the densities of probabilities for the 9 attributes with the most
mean decrease accuracy importance.

On this Figure, each plot, but the PageRank plots, highlight significant
behavior differences between the promoted and the non-promoted candidates:
the interdecile range of the density of probabilities of the promoted
candidates is smaller than the non-promoted candidates one (the curve
is flatter for non-promoted ones). This suggests that the promoted
candidates behave more similarly than the non-ones. This is an explanation
of the better prediction of promotion than of no-promotion in Model
4. Moreover, we can see in the first 6 plots that the peak of density
for the promoted candidates has bigger value than those for the non-promoted:
the successful candidates are more active than the unsuccessful ones.

\begin{figure}
\centering{}three\includegraphics[scale=0.15]{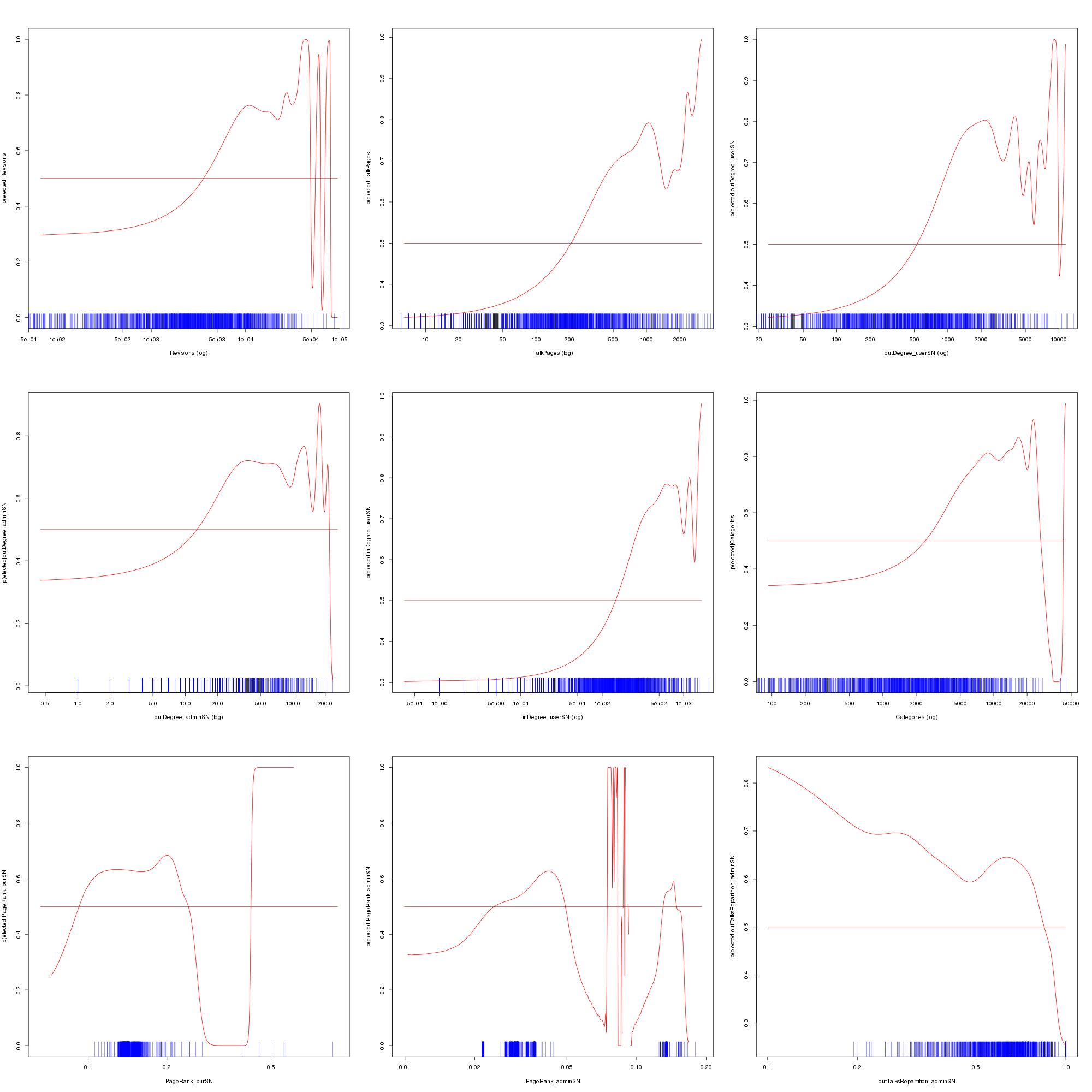}\protect\protect\caption{\label{fig:estimated-probabilities}Estimated probabilities of being
promoted according to the top 9 attributes }
\end{figure}

To estimate these differences, we computed the estimated probability
of being promoted knowing only each attribute, one by one, for the
top 9 attributes (Figure \ref{fig:estimated-probabilities} presents
the results).

The estimated probabilities of being elected according to the number
of revisions ($Revisions$) indicates that the probability of being
promoted exceeds 50\% beyond about 2,500 revisions and is about 70\%
beyond 6000 revisions%
\footnote{It reaches extreme values (0\%, 100\%) beyond 40000 revisions because
there are not enough case studies of candidates with such big number
of revisions. We cannot deduce preferred values on the revision number
based on these extreme cases. However, they show that even with a
huge number of revisions, a candidate can be rejected.%
}.

There is a similar behavior pattern (probability of 50\% beyond a
threshold (T) and extreme values with not enough case studies) for
the probabilities of being promoted according to the following attributes:
$\mathit{Categories}$ ($T{\approx}1700$), $\mathit{TalkPages}$
($T{\approx}130$), $\mathit{outDegree}_{\mathit{userSN}}$ ($T{\approx}450$),
$\mathit{outDegree}_{\mathit{adminSN}}$ ($T{\approx}17$), $\mathit{inDegree}_{\mathit{userSN}}$
($T{\approx}140$).

The attribute $\mathit{outTalksRepartition}_{\mathit{adminSN}}$ shows
that too dispersed candidates in terms of number of talks to admins
(thus in the adminSN graph) reduce their chance to be promoted.

No explicit trend is emerging from the probability according to the
PageRank attributes. We cannot advise a preferred user social networking
behavior with administrator or bureaucrats based on this information

\section{Discussion}

Our results are consistent with the Guide to RfA, with previous results
and with the theories on epistemic communities. Regarding the guide,
we provide much more precise figures of how many contributions and
interactions are needed to have a good probability to be elected,
and we show that there are quite narrow windows in terms of number
of contributions and discussions which maximize the chances. Simplifying
the measures proposed by \citet{BurkeKraut08b} regarding the edit
activities, and adding the activity on social networks leads to a
better evaluation of the chances to be elected (75,6\% of good prediction
for their model, 78\% for our model), keeping the number of explanatory
variables reasonably low.

This said, and as pointed out by \citeauthor{Zhuetal11}'s study (2011)
on the differences between administrative persons (''admin'' or
''sysop'') and project leaders, in the English Wikipedia, who showed
that if local project leaders leave more task oriented messages when
administrators are more in the social exchange, sending more personal
messages in users' personal pages (p. 4): the social activity in the
whole project (Talk and userSN variables) is also very important to
become administrator. We also refined \citet{AntinCheshireNov12}'s
findings that people involved from the beginning in more diverse revision
activities are more likely to take administrative responsibilities.
To be elected, a candidate has to be involved in various talkpages,
in various articles, but not too much! Being too dispersed (not focused)
in terms of of revisions or being excessive in the number of contacts
(too many) lead to a failure.

Regarding the general discussion about how communities of creation
work and can be connected with the study of \citet{RullaniHaefliger13}
on how core developers are recruited in open source software communities.
As supposed for an epistemic community, the contribution in knowledge
(Revision) is the first criterion to be considered as a good candidate,
and most of the non-elected are not because of a lack of contribution.
But, once the candidates have proven their competency (production
of knowledge) and their willing to do the job (interacting with people),
and even when the choice is opened to no-core members as it is in
Wikipedia, knowing and be known by these core members, the future
pees, makes the difference. In our case, the variable $\mathit{outDegree}_{\mathit{adminSN}}$
comes fifth in the most important explanatory variables for a positive
election, and $\mathit{inDegree}_{\mathit{burSN}}$ seventh.

There are obvious limitations to our work. First, our survey addresses
only one project (English Wikipédia) and should be extended to other
languages and other epistemic communities. However, as we already
pointed out, they are consistent with those found in open source software
communities. Second, if our model is good at forecasting the election
(more than 80\%), it is less good for the non-election (around 70\%).
Dropping the extreme cases, the people who are not elected because
they talked too much, maybe because they fought with the administrators,
for instance, may improve the prediction.

But as it is, it already has very practical consequence for the managers
and the people involved in those communities. Promoting and encouraging
people to take responsibilities in voluntary organization is a major
issue for the convenors of communities, being online or offline. Our
results suggest that much precise criterion could be posted on what
is expected from the candidates to administrative functions in those
knowledge production-oriented communities (epistemic communities).

\section{Conclusion}

In this article, for predicting the success in being promoted in communities
of creation, we proposed to consider the social behavior of those
candidates in addition to the usual knowledge behavior, looking at
their interaction with the participants and with their future peers.
We did so looking at the election of administrators in the English
Wikipedia.

We compared three predictive models using the random forest algorithm,
a revision-based model, a social-based model and mixed model based
on social and revisions. We show that using only on social behavior
of candidates, it is possible to predict the promotion results with
a 74.2\% accuracy whereas it is 74.0\% considering only their revision
behavior. Combining social and revision behaviors, we obtain a 77.8\%
accuracy which is a little better than the 75.6\% accuracy given in
\citet{BurkeKraut08b}.

Beyond the predictive model, this article provided estimated probabilities
of being elected according to each attribute. They highlighted thresholds
under which the candidates reduce their chance to be promoted, but
also show that too active candidates in terms of contribution or social
interaction may also find it difficult to be elected. Finally, another
interesting result is that candidates being too dispersed (not focused)
in terms of revisions (many pages with few revisions) and also in
terms of social talks (few talks with many users/admins) reduce their
chance to be promoted.

In other words, and even if our results must be confirmed in other
online epistemic communities, the candidates for responsabilities
in those communities must be aware that beyond the ``professional''
skills requested to be considered for such promotion, taking responsibilities
in those big communities mean working in a team, and that the social
skills, the knowledge of the incumbents, matter too, as social interaction
and coordination are key for the team to be effective, and thus efficient.

 \bibliographystyle{enicolas}
\bibliography{NJ_KR}

\end{document}